\documentclass[fleqn,usenatbib]{mnras}
 
\usepackage[T1]{fontenc}

\DeclareRobustCommand{\VAN}[3]{#2}
\let\VANthebibliography\thebibliography
\def\thebibliography{\DeclareRobustCommand{\VAN}[3]{##3}\VANthebibliography}

\usepackage{graphicx}	
\usepackage{amsmath}	
\usepackage{amssymb}	

\usepackage{lipsum}

\newcommand{\msol}{{M$_\odot$}}

\newcommand{\Mcrit}{$\,M_{200}^{\rm crit}$}

\newcommand{\EAGN}{$E_{\rm AGN}$}

\newcommand{\organic}{{\sc organic}}
\newcommand{\secular}{{\sc secular}}
\newcommand{\merger}{{\sc merger}}
\newcommand{\earlysecular}{{\sc early-secular}}
\newcommand{\earlymerger}{{\sc early-merger}}

\usepackage{newtxtext,newtxmath}

\title[The origin of diversity in SMBH and galaxy growth]{Are the fates of supermassive black holes and galaxies determined by individual mergers, or by the properties of their host haloes?}

\author[J. J. Davies et al.]{Jonathan  J. Davies,$^{1}$\thanks{E-mail: astrojdavies@gmail.com}
Andrew Pontzen$^{1}$\thanks{E-mail: a.pontzen@ucl.ac.uk}
and Robert A. Crain,$^{2}$
\\
$^{1}$Department of Physics and Astronomy, University College London, Gower Street, London WC1E 6BT, UK\\
$^{2}$Astrophysics Research Institute, Liverpool John Moores University, 146 Brownlow Hill, Liverpool L3 5RF, UK
}

\date{Accepted XXX. Received YYY; in original form ZZZ}

\pubyear{2023}

\begin{document}
\label{firstpage}
\pagerange{\pageref{firstpage}--\pageref{lastpage}}
\maketitle
\begin{abstract}
The fates of massive galaxies are tied to the evolution of their central supermassive black holes (BHs), due to the influence of AGN feedback. Correlations within simulated galaxy populations suggest that the masses of BHs are governed by properties of their host dark matter haloes, such as the binding energy and assembly time, at a given halo mass. However, the full picture must be more complex as galaxy mergers have also been shown to influence the growth of BHs and the impact of AGN. In this study, we investigate this problem through a controlled experiment, using the genetic modification technique to adjust the assembly history of a Milky Way-like galaxy simulated with the EAGLE model. We change the halo assembly time (and hence the binding energy) in the absence of any disruptive merger events, and find little change in the integrated growth of the BH. We attribute this to the angular momentum support provided by a galaxy disc, which reduces the inflow of gas towards the BH and effectively decouples the BH's growth from the halo's properties. Introducing major mergers into the assembly history disrupts the disc, causing the BH to grow $\approx 4\times$ more massive and inject feedback that reduces the halo baryon fraction by a factor of $\approx 2$ and quenches star formation. Merger events appear essential to the diversity in BH masses in EAGLE, and we also show that they increase the halo binding energy; correlations between these quantities may therefore be the result of merger events.

\end{abstract}




\section{Introduction}
\label{sec:intro}

Feedback from active galactic nuclei (AGN) is a near-ubiquitous ingredient of modern galaxy formation models, responsible for regulating the growth of galaxies at and above the mass of the Milky Way, quenching star formation in massive galaxies, and maintaining quiescence in the central galaxies of group and cluster haloes by suppressing cooling flows \citep[e.g.][]{bower06,croton06,sijacki07,somerville08,vogelsberger14,schaye15,dubois16,mccarthy17,kaviraj17,tremmel17,henden18,weinberger18,dave19}.

Cosmological simulations predict that AGN feedback has a minimal impact on lower-mass galaxies, as outflows associated with star formation are able to remove gas from the centre of the galaxy and prevent the growth of the supermassive black hole (SMBH). However, above a critical halo mass scale corresponding to that of $L^\star$ galaxies, the entropy of the shock-heated circumgalactic medium (CGM) exceeds that of these outflows; this confines gas to the galaxy centre and allows the SMBH to grow and begin influencing the galaxy-halo ecosystem \citep{dubois15,bower17,mcalpine18,keller20,habouzit20,truong21}.

Above this critical mass scale, the EAGLE, IllustrisTNG and SIMBA simulations exhibit diversity in the properties of SMBHs, galaxies and their CGM. Haloes in which AGN feedback has had little impact tend to be gas-rich and host star-forming central galaxies, whereas haloes hosting overmassive SMBHs that have injected a lot of AGN feedback energy tend to be gas-poor and host quenched galaxies \citep{davies19,davies20,dave19,terrazas20,appleby21,robsondave21,sorini22}. Understanding why the impact of AGN feedback varies in haloes of a given mass is therefore key to understanding why diversity exists in the properties of the $\sim L^{\star}$ galaxies in these simulations.

One possible origin of this diversity could lie in differences in the underlying binding energies (and/or concentrations) of dark matter haloes \citep{boothschaye10,boothschaye11}. This idea can be understood as a consequence of self-regulation; in a more tightly-bound halo, a SMBH must grow more massive and inject more AGN feedback energy to expel gas from the halo centre. In the EAGLE and IllustrisTNG simulations, haloes with higher binding energies tend to host more massive SMBHs, providing evidence for this connection \citep{davies19,davies20}. The binding energy of a halo, in turn, is assumed to be set by its assembly time \citep[e.g.][]{neto07}, a characteristic that is determined by the halo's initial conditions.

On the other hand, observational evidence is emerging for a connection between AGN feedback and galaxy mergers \citep[e.g.][and references therein]{ellison11,ellison19,satyapal14}, and simulations have long predicted that such events can enhance the growth of SMBHs and the impact of AGN feedback \citep{dimatteospringelhernquist05,hopkins06,hopkins10a,sijacki07,sijacki15,bellovary13,dubois15,pontzen17,steinborn18,mcalpine20}; this merger-induced feedback could explain recent observations of a quenching excess in post-merger galaxies \citep{ellison22}. Recently, \citet{davies22} used a controlled galaxy formation experiment to show that differences in the stellar mass ratio (and hence the disruptive influence) of a single merger can have dramatic effects on the growth of the SMBH at the centre of a galaxy, transforming the baryon cycle, the properties of the CGM, and the star formation activity in the central galaxy.

These lines of evidence suggest that the properties of galaxy-CGM ecosystems depend on both the overall assembly time of the host halo and on individual disruptive events that occur throughout the system's assembly. Identifying the relative importance of these factors is challenging, as they are likely to be degenerate; haloes that assemble early tend to reside in more densely clustered environments \citep[e.g.][]{sheth04,gao05,wechslertinker18}, and may therefore assemble by undergoing many disruptive mergers, while later-assembling haloes may have comparatively quiet histories. 

In this study, we perform a controlled galaxy formation experiment using the genetic modification technique \citep{roth16} to independently assess the role of each of these factors, and unveil how the assembly history of a dark matter halo is connected to the evolution of its central galaxy.

\section{Methods}
\label{sec:methods}

For this study, we have performed a suite of simulations using the EAGLE version of the gravity and smoothed-particle hydrodynamics code {\sc gadget3} \citep[last described by][]{springel05}. Using the `zoom' technique \citep[e.g.][]{katz93,bertschinger01} we simulate the evolution of an individual galaxy and its local environment at high resolution, whilst also following the large-scale forces acting on the system by simulating its wider environment with a low-resolution periodic volume. In this section, we explain the selection of our candidate galaxy (Section \ref{sec:methods:ics}), outline how we modify its initial conditions to adjust its assembly history (Section \ref{sec:methods:gm}), and describe how we identify and characterise galaxies and haloes within our simulations (Section \ref{sec:methods:galaxies}).

We also utilise the flagship EAGLE simulation volume (Ref-L100N1504) in Section \ref{sec:results:correlations} to place our genetically-modified galaxies into the context of the wider population. For detailed descriptions of this simulation, we refer the reader to the EAGLE reference articles \citep{schaye15,crain15,mcalpine16}.

\subsection{Construction and evolution of initial conditions}
\label{sec:methods:ics}

We selected our fiducial galaxy from a periodic simulation volume evolved with the Reference EAGLE simulation model \citep{schaye15,crain15} from uniform-resolution initial conditions (ICs) generated by {\sc genetIC} \citep{stopyra20}. We note therefore that the galaxy was not selected from any of the publicly-available EAGLE simulations; selecting a galaxy evolved from ICs created by {\sc genetIC} simplifies the subsequent genetic modification of these ICs. This simulation is 50 comoving Mpc on a side, containing 512$^3$ dark matter particles of mass $3.19\times 10^7$ M$_\odot$ and an initially equal number of baryonic particles of mass $5.6\times 10^6$ M$_\odot$ (a similar mass resolution to that of the flagship EAGLE simulations) and adopts the \citet{planck16} cosmological parameters.

From this simulation, we selected a present-day star-forming disc galaxy of stellar mass $M_\star=4.3\times 10^{10}$ M$_\odot$, the central galaxy of a halo of mass $M_{200}=3.4\times 10^{12}$ M$_\odot$. We selected this galaxy because it lies on the star-forming main sequence (sSFR$=10^{-10.2}$ yr$^{-1}$) and has a CGM mass fraction\footnote{We define $f_{\rm CGM}\equiv M_{\rm CGM}/M_{200}$, where $M_{\rm CGM}$ is the total mass of all gas within the virial radius that is not star-forming.} $f_{\rm CGM} =0.31f_{\rm b}^{\rm cosmic}$ (where $f_{\rm b}^{\rm cosmic}$ is the cosmic baryon fraction, $\Omega_{\rm b}/\Omega_0$) that is close to the present-day median $f_{\rm CGM}$ at this halo mass in the largest EAGLE volume \citep{davies19}. This galaxy is ideal for our purposes, as it resides in a halo of a mass that far exceeds the critical mass, \Mcrit{}, above which BHs are able to grow efficiently in the EAGLE model \citep{bower17,mcalpine18}, and has a simple merger history after the halo exceeds this mass, with only one minor merger of stellar mass ratio\footnote{We define the merger stellar mass ratio $\mu\equiv M_\star^{\rm infall}/M_\star$, where $M_\star^{\rm infall}$ and $M_\star$ are the stellar masses of the infalling and primary galaxy respectively. These masses are measured at the final snapshot output in which the merging subhaloes are clearly distinguishable by our halo finder, {\sc subfind}.} $\mu=0.17$ occurring at $z\approx 0.74$. We will henceforth refer to this galaxy and its halo as our \organic{} system. 

We generate zoomed initial conditions for this system by selecting all particles within three virial radii\footnote{We define the virial radius, $r_{200}$, as the radius of a sphere enclosing 200 times the critical density of the Universe.} of the galaxy (at $z=0$) and identifying the Lagrangian region defined by these particles in the ICs (at $z=99$). We then refine this region with a factor of 27 more particles, and downsample the simulation volume outside this region by a factor of 8, yielding particle masses of $m_{\rm gas}=2.19\times 10^5$ \msol{}, $m_{\rm dm}=1.18\times 10^6$ \msol{}, and $m_{\rm lr}=3.02\times 10^8$ \msol{} for gas, dark matter and low-resolution particles respectively.

We evolve these initial conditions with the EAGLE model, adopting the recalibrated (Recal) parameter values for the subgrid physics as defined by \citet{schaye15} as these were calibrated for a near-identical mass resolution to that of our initial conditions ($m_{\rm gas}=2.26\times 10^5$ \msol{}, $m_{\rm dm}=1.21\times 10^6$ \msol{}). The details of this model and its calibration may be found in the EAGLE reference articles \citep{schaye15,crain15} and for brevity we do not repeat them here. However, it is important to note that models such as EAGLE include stochastic elements; processes such as the conversion of gas particles to star particles and the injection of feedback energy are governed by the drawing of quasi-random numbers that are compared to probabilities set by the properties of the gas \citep[see][]{schayedv08,dvschaye12}. This stochasticity can cause significant uncertainty in the properties of individual systems \citep[see e.g.][]{genel19,keller19,davies21,davies22,borrow22}. Since the zoom simulations in this study are relatively inexpensive to perform, we simulate each set of initial conditions in our experiment with nine random number seeds each to quantify this uncertainty.

\subsection{Producing genetically-modified galaxies}
\label{sec:methods:gm}

To adjust the assembly history of the \organic{} galaxy, we use the genetic modification (GM) technique \citep{roth16,pontzen17} and the {\sc genetIC} software to generate modified sets of ICs for the system. From the overdensity field in the original ICs, {\sc genetIC} finds the closest possible field that also satisfies certain constraints, which we design to produce our desired changes to the halo assembly history. This technique preserves the large-scale environment of the system, and the modified fields remain consistent with a $\Lambda$ cold dark matter ($\Lambda$CDM) cosmology. To assess the role of individual merger events and the overall assembly history of the system independently of each other, we generate four complementary sets of modified ICs using this method. 

First, to examine the influence of merger events at a fixed assembly time, we produce a pair of modified ICs, \secular{} and \merger{}, designed to decrease or increase the stellar mass ratio of the \organic{} system's $z\approx 0.74$ minor merger respectively. This is achieved by identifying the particles bound to the infalling halo at an earlier time ($z=1.73$)\footnote{We choose this time as it corresponds to the final snapshot output in which the merging haloes are clearly distinguishable by the {\sc subfind} algorithm.}, tracing these particles back to their locations in the ICs, and decreasing or increasing the mean overdensity, $\bar{\delta}$, in the patch of the field defined by these locations. To preserve the overall mass accretion history we also apply two further constraints; $\bar{\delta}$ in the patch defined by particles that comprise the main halo at $z=1.73$ is kept fixed, as is $\bar{\delta}$ in the patch corresponding to the $z=0$ halo to ensure that the same final halo mass is reached.

We also test the influence of the overall assembly time independently of merger events, by comparing the evolution of galaxies that have differing assembly times and experience no significant mergers that would be able to drive black hole (BH) growth. To do so, we compare the \secular{} system with another modified variant of our galaxy, which assembles earlier and experiences no significant mergers after the \Mcrit{} threshold is reached. We produce this behaviour by assembling more mass into the main progenitor at early times ($z=2$) whilst keeping the final halo mass fixed, which has the effect of both accelerating the halo assembly and reducing the significance of all $z<2$ mergers. This is achieved in practice by increasing $\bar{\delta}$ in the patch of the ICs corresponding to the $z=2$ halo, while keeping $\bar{\delta}$ fixed within the patch corresponding to the $z=0$ halo. We refer to the system evolved from these conditions as the \earlysecular{} system.

Finally, so that we can assess how the \earlysecular{} system would evolve if it had a more disruptive evolution after exceeding \Mcrit{}, we further modify the \earlysecular{} ICs to increase $\bar{\delta}$ in a patch corresponding to an infalling system at $z=3$ with the aim of inducing a subsequent major merger of similar mass ratio to that in the \merger{} system. We refer to the system evolved from these conditions as the \earlymerger{} system.

\subsection{Identifying and characterising galaxies and their haloes}
\label{sec:methods:galaxies}

Haloes are identified on-the-fly in our simulations by applying the friends-of-friends (FoF) algorithm to the dark matter distribution, with a linking length of 0.2 times the mean interparticle separation. Gas, star and BH particles are then assigned to the FoF halo of their nearest dark matter particle. In post-processing, we then identify bound haloes using the {\sc subfind} algorithm \citep{springel01,dolag09}, and use the analysis packages {\sc pynbody} \citep{pontzen13} and {\sc tangos} \citep{pontzentremmel18} to calculate the properties of galaxies and their haloes.

We use {\sc tangos} to construct merger trees that link haloes to their progenitors and descendants based on the number of particles they have in common. Starting with our \organic{} system, we identify the main branch of the tree by calculating the sum of the stellar masses along each possible branch, and then selecting the branch with the largest value. We then identify the \organic{} system's counterparts in other simulations by matching on the number of particles in common at $z=8$, and then tracking this system forwards in time. This yields a more stable and reliable matching between simulations than attempting to find the \organic{} system's counterpart at each output time separately.

We calculate the properties of haloes, such as the halo mass ($M_{200}$) and baryon fraction ($f_{\rm b}\equiv M_{\rm b}/M_{200}$, where $M_{\rm b}$ is the total mass in baryons) within one virial radius of the halo centre of mass. We find the centre of mass with {\sc pynbody}, using the shrinking-sphere method \citep{power03}. The properties of galaxies, such as the stellar mass ($M_\star$) and specific star formation rate (sSFR) are calculated within a spherical aperture of radius 30 physical kpc about the centre of mass. 

We calculate the intrinsic inner-halo binding energy ($E_{2500}^{\rm DMO}$, i.e. that which is set by the halo's assembly history and initial conditions, and not by dissipative baryonic processes\footnote{For further explanation of why we use binding energies from a dark-matter-only simulation, see Section 3 of \citet{davies19}.}) by matching each system to its counterpart in an equivalent dark matter-only simulation and summing the binding energies of all particles within a radius enclosing 2500 times the critical density. For our zoom simulations, we perform this matching using {\sc tangos}, and for the large-volume Ref-L100N1504 simulation we use the bijective particle matching algorithm described by \citet{schaller15}.

We calculate the total energy injected through stellar feedback, $E_\star$, by summing the energies contributed by all star particles within a 30 pkpc aperture about the centre of mass; when a gas particle $i$ is converted into a star particle it provides an energy given by
\begin{equation}
    E_{\star,i} = 1.74 \times 10^{49}\, \mathrm{erg}\, \left(\frac{m_{\star,i}^{\rm init}}{1\,\mathrm{M}_\odot}\right)\, f_{{\rm th},i}(n_{{\rm H},i},Z_i),
\end{equation}
where $m_{\star,i}^{\rm init}$ is the initial stellar mass and $f_{{\rm th},i}$ is an efficiency that depends on the density $n_{{\rm H},i}$ and metallicity $Z_i$ of the gas particle at the time of conversion \citep[for more information see][Section 4.5]{schaye15}. The total feedback energy injected through AGN feedback by the galaxy's central BH is given by
\begin{equation}
    E_{\rm AGN} = \frac{\epsilon_{\rm f}\epsilon_{\rm r}}{1-\epsilon_{\rm r}}(M_{\rm BH}-M_{\rm BH}^{\rm seed}) c^2,
\end{equation}
where $M_{\rm BH}$ is the BH mass, $c$ is the speed of light, $\epsilon_{\rm r}=0.1$ is the radiative efficiency of the accretion disc and $\epsilon_{\rm f}=0.15$ is the fraction of the radiated energy that couples to the surrounding gas. $M_{\rm BH}^{\rm seed}=10^5$ \msol{}$/h$ is the mass of the BH seed, which does not contribute any feedback energy. We note that these definitions include ``ex-situ" energy injected by stars and BHs in other progenitor galaxies that subsequently merged with our main system, but since this energy directly impacts the properties of these progenitors we include it in our definition.

\section{Results}
\label{sec:results}

We first examine the impact of a varying halo assembly time in the absence of merger events in Section \ref{sec:results:assembly}, before exploring how our systems evolve when mergers do occur in Section \ref{sec:results:mergers}. Finally, in light of our findings, we discuss the origin of previously-identified correlations in the EAGLE galaxy population in Section \ref{sec:results:correlations}. 

\subsection{The impact of halo assembly time in the absence of merger events}
\label{sec:results:assembly}

We begin by isolating the effect of a varying assembly time on the evolution of our system. We do so by comparing the evolution of the \secular{} and \earlysecular{} haloes, which have different assembly times and experience no significant mergers after their masses exceed \Mcrit{}, the threshold above which disruptive events may lead to BH growth.

\subsubsection{Halo and galaxy evolution}

\begin{figure}
\includegraphics[width=\columnwidth]{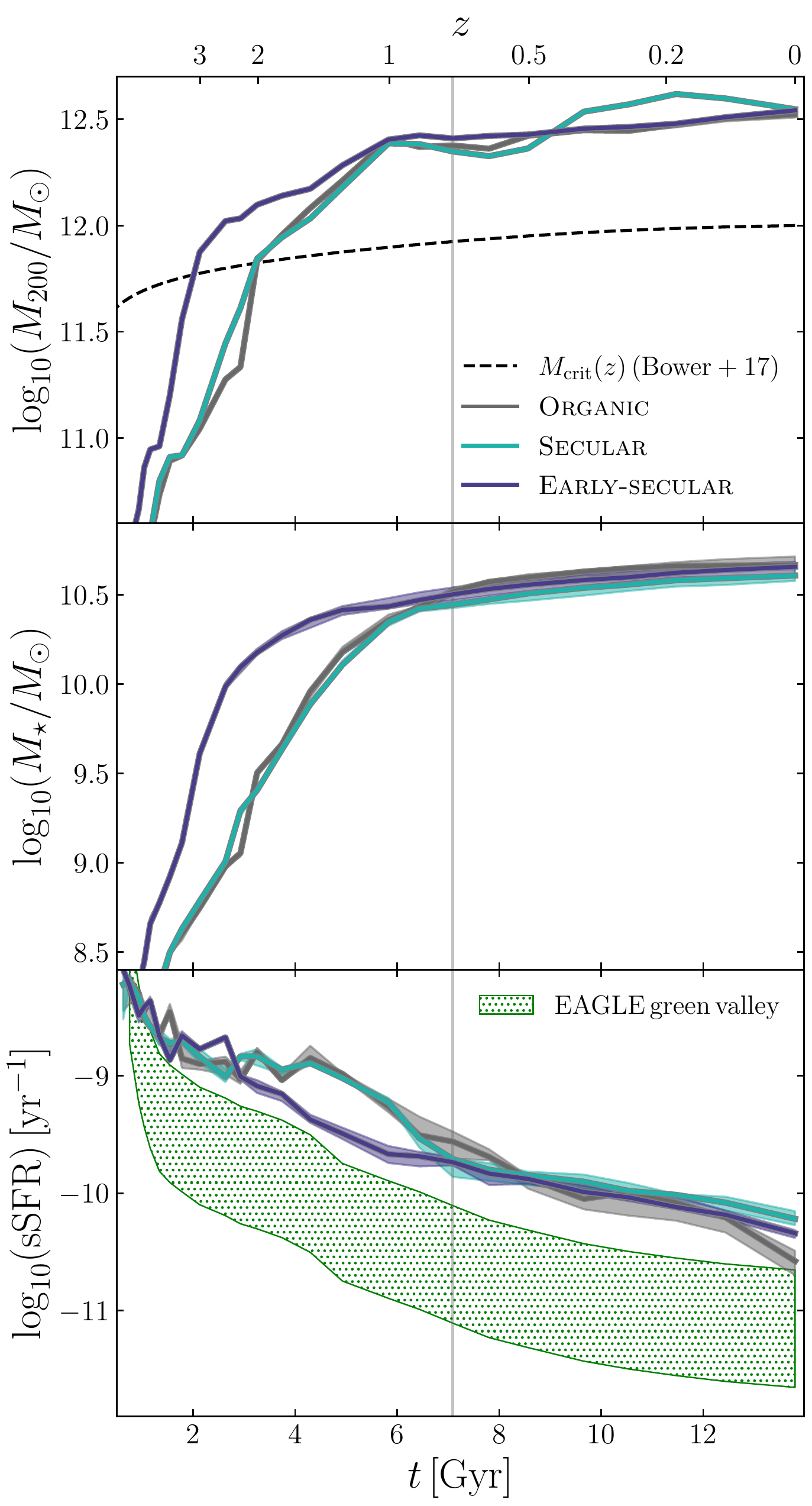}
\vspace{-4mm}
\caption{The \organic{} system and our modified \secular{} and \earlysecular{} variants reach approximately the same final halo mass ($M_{200}$) and stellar mass ($M_\star$), but exhibit differences in their mass histories due to our genetic modifications. All three systems remain star-forming until $z=0$, despite their halo masses being far in excess of \Mcrit{}. The upper, middle and lower panels show the evolution of $M_{200}$, $M_\star$ and the specific star formation rate (sSFR) respectively for these three systems. Solid lines show the median values for nine resimulations of each set of initial conditions with different random number seeds, and shading indicates the interquartile range. The time of the first snapshot output following any significant merger (stellar mass ratio $\mu>0.1$) is shown with a vertical line. We indicate the critical halo mass, \Mcrit{}, above which black holes can grow efficiently in EAGLE with a dashed black line in the top panel. The location of the green valley in the EAGLE population is shown with green hatching in the lower panel.}
\vspace{-4mm}
\label{fig:secularmass}
\end{figure}

Figure \ref{fig:secularmass} shows how the \organic{} system and its \secular{} and \earlysecular{} variants evolve in terms of their halo mass ($M_{200}$, upper panel), stellar mass ($M_\star$, middle panel) and specific star formation rate (sSFR$\equiv$SFR$/M_\star$, integrated over the preceding 100 Myr\footnote{Our results are not strongly sensitive to this choice, and we have verified that using longer timescales of 300 Myr and 1 Gyr yield similar results.}, lower panel). In this and other figures of this type, we indicate the first output after which a significant merger has occurred with vertical lines, coloured to match the simulation in which the merger occurred. We define significant mergers to be those in which the stellar mass ratio, $\mu$, is greater than 0.1. We quantify the uncertainty that arises from EAGLE's stochastic subgrid physics implementation by simulating nine realisations of each set of ICs, varying the seed used for the quasi-random number generator each time. Solid lines indicate the median value at each output time, and shading shows the interval between the third and seventh values in the rank-ordered distribution, a good approximation of the interquartile range (IQR).

The halo mass histories of the \organic{} and \secular{} galaxies are very similar. We indicate the evolution of \Mcrit{} with a dashed line; both systems exceed this threshold at approximately the same time ($t\approx 3.2$ Gyr). The \earlysecular{} halo assembles more quickly, and exceeds \Mcrit{} earlier, at $t\approx 2$ Gyr. As intended, all systems converge to near-identical halo masses of $10^{12.52}$ \msol{} (\organic{}, IQR$=0.02$ dex) and $10^{12.54}$ \msol{} (\secular{} and \earlysecular{}, IQR$=0.02$ and 0.01 dex respectively). The \secular{} halo temporarily has a higher mass than the others at $z\approx 0.4-0.1$; this is due to mass bound to another halo passing within $r_{200}$ of the main halo during this period. This halo passes closer to the \secular{} system than it does to the \organic{} and \earlysecular{} systems as an unintended side-effect of our modifications; at its closest approach, its centre of mass lies approximately at the virial radius of the main halo.

As a result of its earlier assembly time, the \earlysecular{} halo is intrinsically more tightly-bound than the \secular{} halo. When simulated with purely collisionless dynamics, the binding energy of the inner halo, $E_{2500}^{\rm DMO}$, is 30\% higher for the \earlysecular{} system; this difference corresponds to $\approx 1\sigma$ in the flagship EAGLE simulation at this halo mass\footnote{We calculated this statistic for haloes in the collisionless (DMONLY) EAGLE L100N1504 simulation, in a 0.1 dex wide mass window about $M_{200}=10^{12.54}$ \msol{}.}.

The stellar mass histories of the \organic{} and \secular{} galaxies also trace each other closely, but diverge slightly at $t=7.1$ Gyr, as a result of the \organic{} galaxy's sole significant merger (after \Mcrit{} is exceeded); a minor merger of stellar mass ratio $\mu=0.17$. The ex-situ stellar mass contributed by this merger elevates the \organic{} galaxy's stellar mass above that of its secularly-evolving counterpart, but by the present day the \secular{} system catches up through predominantly in-situ star formation, and the \organic{} and \secular{} galaxies reach stellar masses of $10^{10.67}$\msol{} (IQR$=0.07$ dex) and $10^{10.61}$\msol{} (IQR$=0.04$ dex) respectively. The \earlysecular{} galaxy attains a similar final stellar mass of $10^{10.66}$\msol{} (IQR$=0.07$ dex), but assembles the majority of that mass earlier as a result of an accelerated halo assembly history. Neither the \secular{} or \earlysecular{} galaxies experiences any significant mergers after the halo mass exceeds \Mcrit{}, and therefore by comparing the properties of these systems we may examine the influence of the overall halo assembly time in the absence of mergers.

In the lower panel of Figure \ref{fig:secularmass} we show how these assembly histories influence the specific star formation rates of our galaxies. We also show the location of the green valley as a function of time for galaxies of a comparable stellar mass in the largest EAGLE simulation volume (Ref-L100N1504). We define this based on the sSFR of all EAGLE galaxies in a 0.2 dex-wide window about the current stellar mass of the \organic{} galaxy. Following \citet{wright19}, we take the locus of the star-forming main sequence to be the mean sSFR of these galaxies, subject to a floor $\log_{10}({\rm sSFR/yr^{-1}})>-11+0.5z$, and define the green valley to lie within 5\% and 50\% of this value.

All three systems remain actively star-forming throughout their evolution. The earlier stellar mass assembly of the \earlysecular{} galaxy is reflected in its sSFR, which is higher at $z\approx 3-2$, and lower at $z\approx 1.5-1$, than the later-assembling systems. However, there is little difference between its final sSFR of $10^{-10.34}$ yr$^{-1}$ (IQR$=0.07$ dex) and the final sSFR of the later-assembling \secular{} system $10^{-10.2}$ yr$^{-1}$, IQR$=0.1$ dex). Therefore, while correlations within the wider EAGLE population indicate that earlier-assembling and more tightly-bound haloes tend to host central galaxies with lower sSFR \citep[e.g.][]{matthee19,davies19,davies20}, we find that adjusting these quantities alone does not causally affect the central galaxy's present-day star formation activity.

\subsubsection{Feedback and the baryon fraction}
\label{sec:secularfb}

\begin{figure}
\includegraphics[width=\columnwidth]{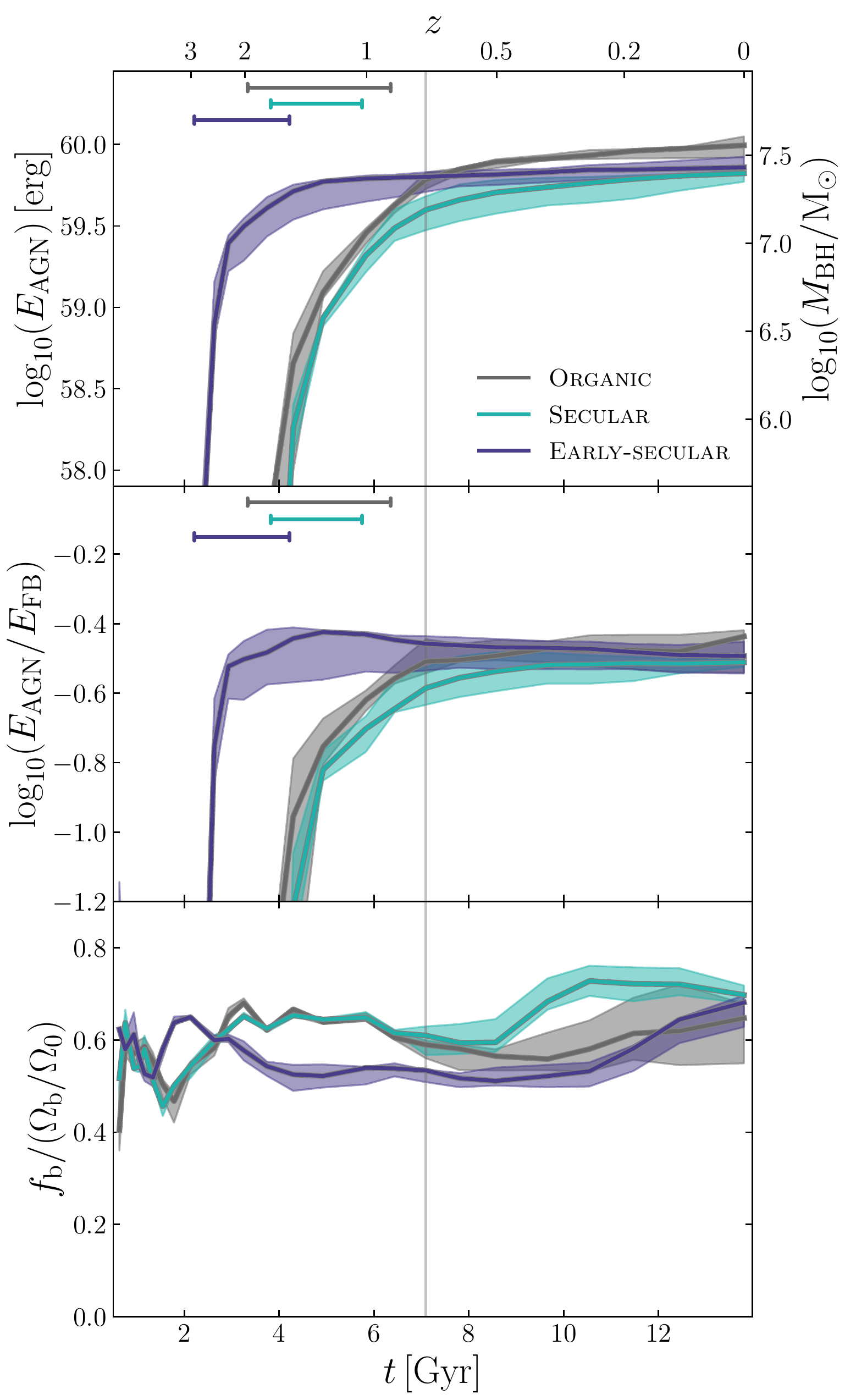}
\vspace{-4mm}
\caption{Differences in assembly history make little difference to $E_{\rm AGN}$ and its contribution to the overall feedback budget in the absence of disruptive events, and all systems remain baryon-rich. In the same fashion as Figure \ref{fig:secularmass}, the upper, middle and lower panels show the evolution of the integrated energy injected by AGN feedback ($E_{\rm AGN}$), the fraction of the total feedback energy, $E_{\rm FB}$, contributed by AGN, and the halo baryon fraction ($f_{\rm b}$, normalised to the cosmic fraction, $\Omega_{\rm b}/\Omega_0$) respectively, for the \organic{}, \secular{} and \earlysecular{} systems. We illustrate the phase of non-linear growth (NLG) undergone by the central black hole in each system with horizontal bars.}
\vspace{-4mm}
\label{fig:secularenergy}
\end{figure}

To investigate why changes in the assembly time alone do not change the present-day star formation rates of our galaxies, we now explore how the feedback history and the halo baryon content are affected by our modifications. In the upper panel of Figure \ref{fig:secularenergy}, we show how the total AGN feedback energy, \EAGN, injected into our systems as a function of time, and in the middle panel we show the fraction of the total feedback energy, $E_{\rm FB}=E_\star+E_{\rm AGN}$, that is contributed by AGN feedback. In the lower panel, we show how this energy injection influences the halo baryon fraction ($f_{\rm b}$, normalised to the cosmic fraction $f_{\rm b}^{\rm cosmic}=\Omega_{\rm b}/\Omega_0$, lower panel). Details of how each of these quantities was calculated are given in Section \ref{sec:methods:galaxies}; since we calculate $E_{\rm AGN}$ directly from the central SMBH mass, $M_{\rm BH}$, we show this mass as a second axis in the upper panel. 

To place the growth of the BHs in our systems in context, we briefly review the general behaviour of BHs in the EAGLE model. \citet{mcalpine18} proposed that EAGLE BHs grow in three phases:
\begin{enumerate}
    \item In lower-mass haloes, stellar feedback is able to efficiently expel gas from the central regions of galaxies, keeping the gas density in the BH's vicinity, $\rho_{\rm BH}$, low. The BH is therefore unable to grow and remains close to the mass at which it was seeded. \citet{mcalpine18} name this the {\it stellar feedback regulated} phase.
    
    \item As $M_{200}$ reaches \Mcrit{} and the halo virial temperature approaches $10^{5.6}$ K, stellar feedback-driven outflows have lower entropy than the CGM and are no longer buoyant. These outflows can no longer remove gas from the galaxy centre, causing $\rho_{\rm BH}$ to increase significantly. EAGLE BHs then accrete gas according to a modified version of the spherically-symmetric \citet{bondi44} formula \citep[see][]{rosasguevara15}, growing at a non-linear rate proportional to $M_{\rm BH}^2$ for as long as $\rho_{\rm BH}$ remains high. \citet{mcalpine18} name this the {\it non-linear growth (NLG)} phase. Following their study, we take the NLG phase to be where d$\log_{10}(M_{\rm BH})/$d$t>0.25$ dex Gyr$^{-1}$ and illustrate it with horizontal bars in Figure \ref{fig:secularenergy}.
    
    \item Once the BH becomes sufficiently massive that its AGN feedback can expel gas from the BH's immediate vicinity, $\rho_{\rm BH}$ falls and the NLG phase ends. Thereafter, the BH can regulate its own growth (and $\rho_{\rm BH}$) through AGN feedback. \citet{mcalpine18} name this the {\it AGN feedback regulated} phase. However, as we will demonstrate in this study, AGN feedback does not necessarily dominate the energy injected into the galaxy and halo after the NLG phase ends.
    
\end{enumerate}

During the NLG phase, \EAGN{} and $M_{\rm BH}$ rise sharply, and this occurs markedly earlier for the \earlysecular{} system than for the later-assembling \organic{} and \secular{} systems. However, over the lifetimes of the \secular{} and \earlysecular{} systems, the total energy injected by AGN feedback is not significantly different ($E_{\rm AGN}=10^{59.8}$ erg with IQR$=0.1$ dex), and is slightly less than the energy injected into the \organic{} system ($E_{\rm AGN}=10^{60.0}$ erg, IQR$=0.1$ dex). 

In the absence of merger events, earlier halo assembly and a 30\% higher binding energy therefore does not necessarily yield a more massive BH or the injection of more AGN feedback energy. This is somewhat surprising, since the model of \citet{boothschaye10,boothschaye11} predicts that BHs will grow until they have injected an energy set by the halo binding energy, and strong positive correlations exist between $M_{\rm BH}$ and $E_{2500}^{\rm DMO}$ at fixed $M_{200}$ in the EAGLE population \citep{davies19}. While the final $M_{\rm BH}$ and \EAGN{} do not change, the \earlysecular{} BH does grow at a faster rate during the NLG phase, and we find that it attains a higher $\rho_{\rm BH}$ during this period. This is consistent with the behaviour of the wider EAGLE population, in which BHs that enter the NLG phase earlier are more luminous and have higher Eddington ratios during the phase \citep{mcalpine18}. This may be the result of a higher cosmological infall rate and/or mean density of the universe at earlier times, or be due to the higher halo binding energy making it harder for feedback processes to reduce the central gas density.

AGN feedback does not dominate the energy input into any of these systems, as shown in the middle panel; it contributes $\approx 30$\% of the feedback injected into the \secular{} and \earlysecular{} systems and $\approx 40$\% of the feedback in the \organic{} system. $E_{\rm AGN}/E_{\rm FB}$ sharply increases as each system's BH undergoes its NLG phase, but then remains fairly constant once this phase is complete; the post-NLG phase therefore need not be dominated by AGN feedback. Both stellar and AGN feedback continue to operate in tandem for the remainder of each system's evolution, apparently co-existing in equilibrium and contributing similar rates of energy injection. 

The lower panel shows that the feedback injected into these systems has not led to a strong expulsion of baryons from their haloes. Prior to each system's NLG phase, the baryon fraction $f_{\rm b}$ is approximately $0.65 f_{\rm b}^{\rm cosmic}$, lower than the cosmic fraction due to the effects of photoionisation and stellar feedback at early times. In all cases, the AGN feedback injected during the NLG phase causes a small decrease in $f_{\rm b}$, and in the \organic{} halo the additional AGN feedback induced by the minor merger causes its $f_{\rm b}$ to fall below that of the \secular{} halo. Overall, the haloes remain baryon rich, retaining approximately $60-70\%$ of the cosmic baryon fraction within their haloes at the present day. 

In EAGLE and other galaxy formation models, quenching galaxies appears to require that a large fraction of the halo's baryons are expelled beyond the virial radius by AGN feedback, in order to reduce the cooling rate of CGM gas onto the interstellar medium \citep{davies20,zinger20}. This does not occur for our secularly-evolving systems, explaining why they remain star-forming until the present day.

\subsubsection{The role of the galaxy disc}

The results discussed so far demonstrate that, in the absence of mergers or other disruptive events, the assembly time and intrinsic binding energy of a dark matter halo do not strongly influence the present-day properties of our galaxy and its halo. This result can be understood by considering the dynamics of the gas in the vicinity of the central BH. As discussed in Section \ref{sec:secularfb}, EAGLE BHs can only grow rapidly while the density of gas in their vicinity ($\rho_{\rm BH}$) remains high. If gas is kept away from the BH, spread out in a co-rotating disc, $\rho_{\rm BH}$ will be lower and the BH's growth will be inhibited.

Using a similar genetic modification experiment, \citet{davies22} showed that suppressing the influence of a merger also suppresses the growth of the central BH, because the co-rotational motion of the gas is preserved in the absence of any disruptive events. A similar phenomenon is occurring for our \secular{} and \earlysecular{} galaxies, which have persistent, strongly co-rotating discs throughout their evolution. Following the process outlined in Section \ref{sec:secularfb}, their BHs undergo an NLG phase when $M_{200}\to$\Mcrit{}, before settling into the AGN feedback-regulated phase. For the remainder of their evolution, 80-90\% of the kinetic energy of the gas within 3 physical kpc of their BHs is invested in co-rotational motion\footnote{These values are equivalent to the $\kappa_{\rm co}$ diagnostic \citep{correa17} for the gas, calculated with the routines of \citet{thob19}.}, and so their BHs never grow rapidly again as the rotational support keeps $\rho_{\rm BH}$ low. Figure \ref{fig:secularenergy} shows that the halo assembly time and binding energy influence {\it when} the NLG phase occurs, and how rapid the BH growth is in that phase, but they do not significantly change the AGN feedback energy required for the BH to reduce $\rho_{\rm BH}$ and regulate its own growth. The final BH mass is therefore similar in each case.

Our findings show that when galaxies are undisturbed by mergers and retain gaseous discs, they settle into a state where a balance of stellar feedback and low-level AGN feedback is sufficient to regulate the growth of the BH and the galaxy, long after the \Mcrit{} threshold has been exceeded. Their BHs inject only the minimum amount of energy required to end the NLG phase, and afterwards only require a small amount of AGN feedback energy to regulate the buildup of gas that migrates to the centre of the disc. This feedback does not strongly affect the CGM, which can continue cooling onto the galaxy, where the majority of it settles onto the disc and forms stars, and a minority fuels slow and steady growth of the BH.

In this state, the growth of the BH and the properties of the host halo are effectively decoupled. The BH does not need to `offset' the cooling of gas from the whole halo in order to regulate its growth, as is the case in more massive group/cluster systems \citep[e.g.][]{mcnamara07,fabian12}; instead, only a small fraction of the gas cooling from the halo can fuel the BH. While a higher binding energy (and deeper potential) may in principle mean that more AGN feedback is required to expel this gas from the centre of the disc, the near-identical \EAGN{} for our secularly-evolving galaxies demonstrates that this is not a significant effect. As a result, the integrated growth of the BH, and hence its influence on the CGM and the galaxy, is not sensitive to changes in the binding energy of the halo.

\subsection{The impact of merger events}
\label{sec:results:mergers}

We now explore how our \secular{} and \earlysecular{} galaxies and their haloes evolve when they do experience a disruptive merger after exceeding \Mcrit{}. We therefore turn to the systems evolved from our \merger{} and \earlymerger{} initial conditions. Each of these systems undergoes a major merger with a stellar mass ratio $\mu=0.46$, occurring at $t=6.4$ Gyr (\merger) and $t=4.9$ Gyr (\earlymerger). While the stellar mass ratios of these mergers are the same, we note that they are not the {\it same merger}, and they differ in their geometry (i.e. impact parameter and infall angle) and dynamics.

\subsubsection{Halo and galaxy evolution}

\begin{figure}
\includegraphics[width=\columnwidth]{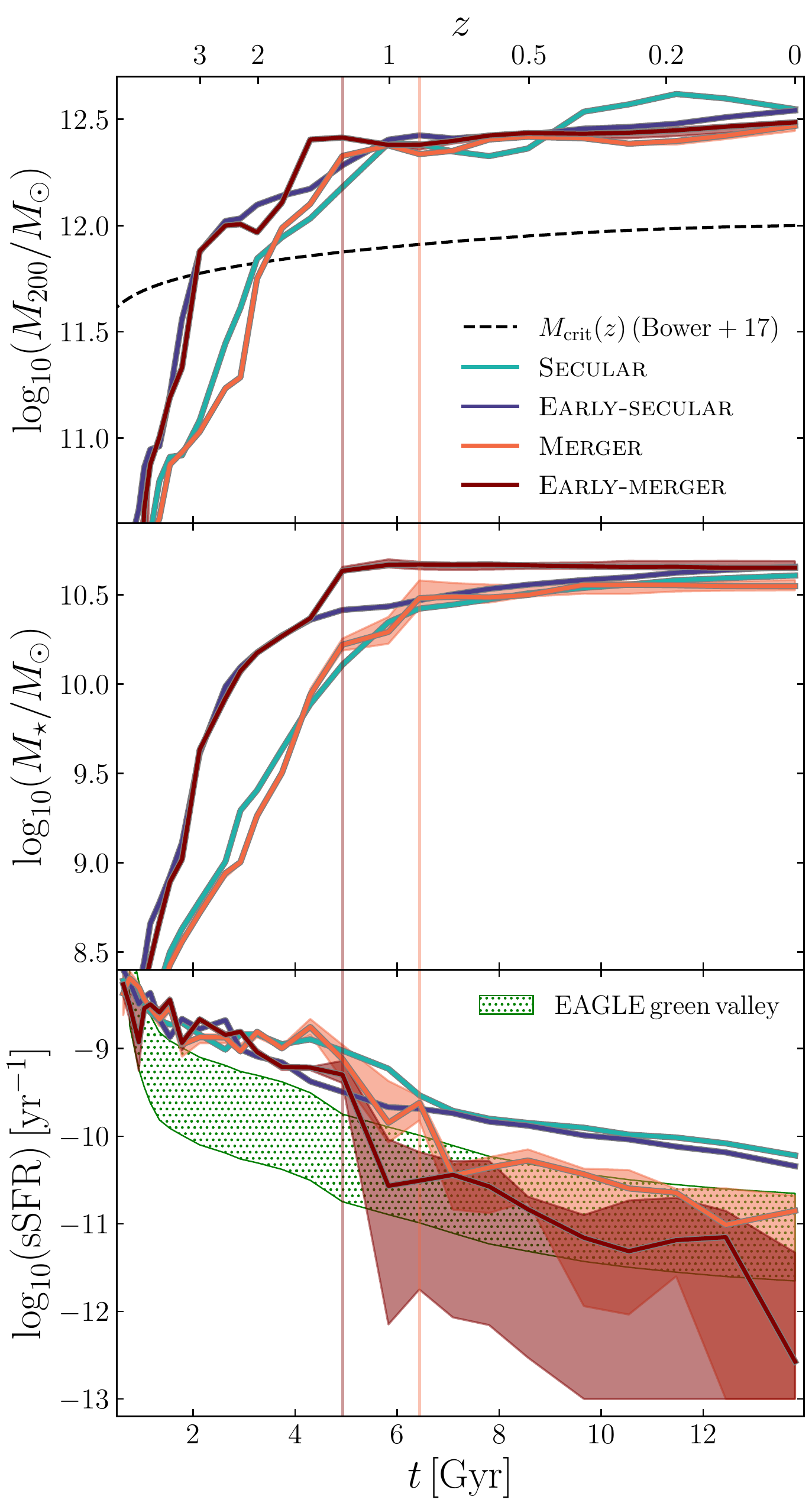}
\vspace{-4mm}
\caption{Our galaxies modified to undergo major mergers attain similar final $M_{200}$ and $M_\star$ to their secularly-evolving counterparts. However, the mergers cause them to leave the main sequence, enter the green valley, and in many cases quench. The three panels are equivalent to those in Figure \ref{fig:secularmass}, but now show the evolution of $M_{200}$, $M_\star$ and the sSFR for the \merger{} and \earlymerger{} systems. The median evolution for the \secular{} and \earlysecular{} systems is shown for comparison.}
\vspace{-4mm}
\label{fig:mergermass}
\end{figure}

Figure \ref{fig:mergermass} is identical in format to Figure \ref{fig:secularmass}, but now focuses on the evolution of the \merger{} and \earlymerger{} halo mass, stellar mass, and sSFR. To demonstrate the differences induced by mergers relative to a secularly-evolving case, we also include the \secular{} and \earlysecular{} systems but only show the median evolution, for clarity. As in previous figures, we highlight the snapshot times after which significant mergers have occurred for the \merger{} and \earlymerger{} systems with vertical lines.

As shown in the upper panel, the halo mass histories of the \secular{} and \merger{} haloes are similar to each other, as are those of the \earlysecular{} and \earlymerger{} haloes, reaching near-identical halo masses by the present day. The later-assembling pair of systems cross the \Mcrit{} threshold at approximately the same time, as do the earlier-assembling pair. The stellar mass histories (middle panel) are also very similar within the earlier- and later-assembling pairs of systems. As with the halo mass, the histories differ at the time of the merger, as the \merger{} and \earlymerger{} galaxies gain a significant ex-situ contribution to their stellar mass, while their secularly-evolving counterparts form stars more steadily in-situ.

The post-merger stellar mass histories of the \merger{} and \earlymerger{} galaxies can be understood by examining the evolution of their sSFR, shown in the lower panel of Figure \ref{fig:mergermass}. Prior to the mergers taking place, the galaxies are on the main sequence, closely following the sSFR of their secularly-evolving counterparts. However, the sSFR declines following the mergers in both cases, with a stronger decline seen for the \earlymerger{} galaxy. The scatter introduced by stochasticity is particularly notable here; in the \earlymerger{} case, many realisations of the galaxy rapidly fall through the green valley and quench within the time interval between simulation outputs, whereas others fall into the green valley and remain there until $z=0$. In general, the median \earlymerger{} galaxy is quenched due to the merger it experiences, whereas the median \merger{} galaxy resides in the green valley.

\subsubsection{Feedback and the baryon fraction}

Figure \ref{fig:mergermass} shows that individual merger events are able to induce strong changes in the sSFR of our galaxy, to a degree that adjusting the overall assembly time alone did not. \citet{davies22} showed that mergers induce such changes because they disrupt the co-rotational motion of gas in the galaxy disc, greatly increasing the density of gas in the central BH's vicinity and allowing the BH to grow more massive (and thus inject more AGN feedback energy) than in a secularly-evolving system. This phenomenon also occurs for our \merger{} and \earlymerger{} systems; prior to the mergers, 80-90\% of the kinetic energy of the gas within 3 physical kpc of their BHs is invested in co-rotational motion, but in the snapshots following their merger events, this fraction has fallen to $\sim 30$\%. This greatly enhances the growth of the BHs in these galaxies.

\begin{figure}
\includegraphics[width=\columnwidth]{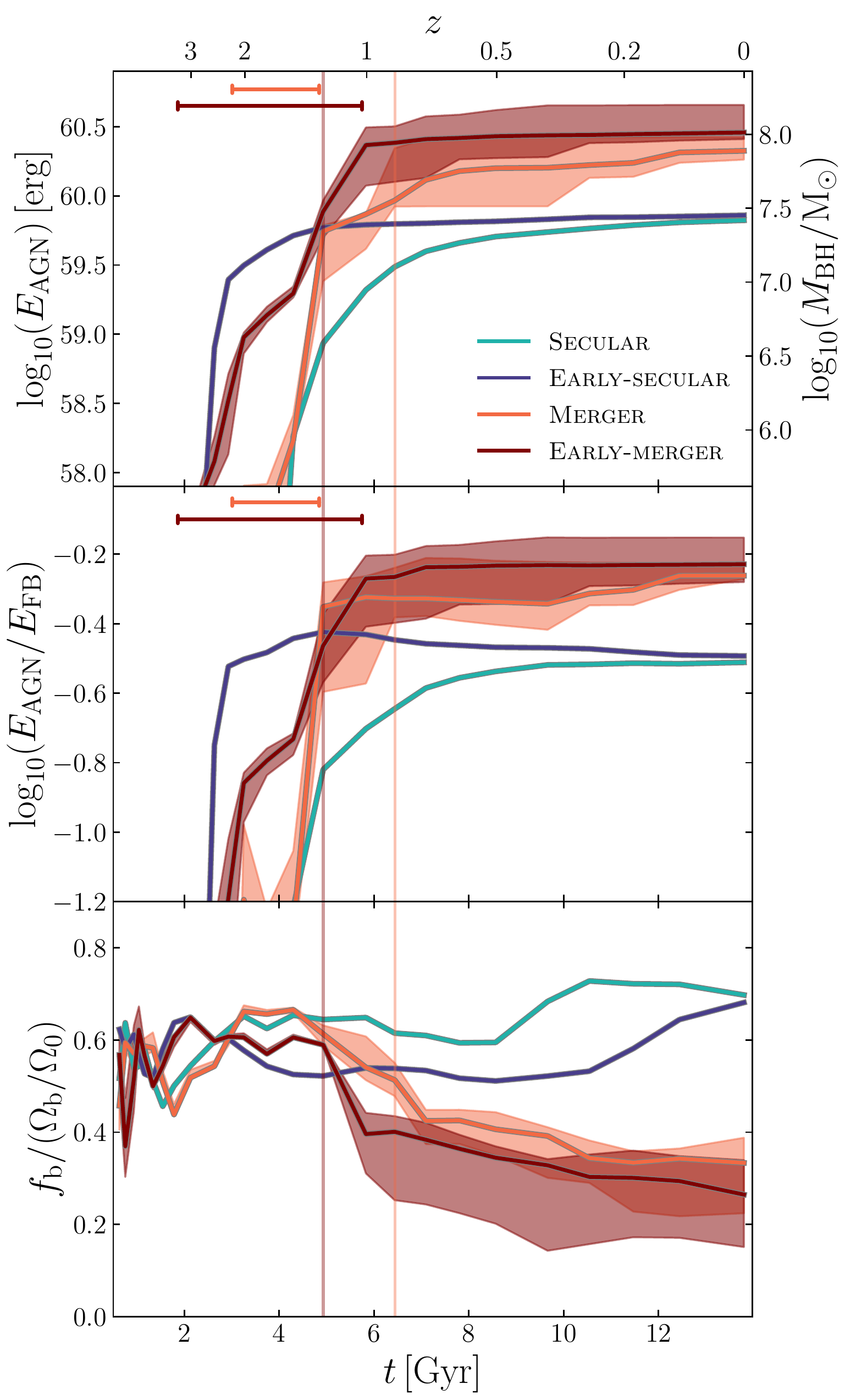}
\vspace{-4mm}
\caption{Mergers cause the injection of significantly more AGN feedback in the \merger{} and \earlymerger{} systems, leading AGN feedback to constitute a greater fraction of the feedback energy injected into these systems. The expulsive effect of this feedback depletes the baryon content of their haloes. The three panels are equivalent to those in Figure \ref{fig:secularenergy}, but now show the evolution of the feedback energetics and halo baryon content for the \merger{} and \earlymerger{} systems, with the median evolution for their secularly-evolving counterparts shown for comparison. We illustrate the phase of non-linear growth (NLG) undergone by the central black hole in each system with horizontal bars.}
\vspace{-4mm}
\label{fig:mergerenergy}
\end{figure}

We show how this disc disruption affects the BH's growth in Figure \ref{fig:mergerenergy}, which has the same format as Figure \ref{fig:secularenergy} but now focuses on the feedback energetics and halo baryon content of the \merger{} and \earlymerger{} systems, again also showing the median evolution of their secularly-evolving counterparts. The haloes exceed \Mcrit{} at approximately the same time as their secularly-evolving counterparts, and so their central BHs begin their NLG phases at similar times. The masses of the BHs then quickly exceed those in the secularly-evolving galaxies once the merger events begin to influence the amount of gas available to them. To follow this process in more detail, we performed additional simulations with 10 Myr output time resolution, finding that BH growth in the \merger{} system is stimulated at the infalling halo's first periapsis over 1 Gyr before the merger, whereas in the \earlymerger{} system it is the final coalescence that triggers rapid BH growth. The final masses of these BHs are $10^{7.9}$ \msol{} (\merger{}, IQR$=0.2$ dex) and $10^{8.0}$ \msol{} (\earlymerger{}, IQR$=0.2$ dex), significantly exceeding the masses of the BHs in the secularly-evolving galaxies ($10^{7.4}$ \msol{}, IQR$=0.1$ dex in both cases), and consequently factors of $\approx 3$ and $\approx 4$ times more AGN feedback energy is injected in these systems respectively.

The middle panel of Figure \ref{fig:mergerenergy} shows that mergers significantly increase the fraction of the total feedback energy that is contributed by AGN feedback; by the present day AGN feedback accounts for $\approx 55$\% of the feedback in the \merger{} system and $\approx 59$\% in the \earlymerger{} system (cf. 30\% in the secularly-evolving systems). This is predominantly the result of a large increase in \EAGN; the mergers increase the integrated stellar feedback energy by only small factors of 1.2 and 1.3 for the \merger{} and \earlymerger{} galaxies respectively. This additional AGN feedback expels a large fraction of the CGM from the \merger{} and \earlymerger{} haloes; as shown in the lower panel they both retain only $\approx 30$\% (IQR$=20$\%) of the cosmic baryon fraction within their haloes at the present day. This reduction in the baryon content of the CGM causes it to cool less efficiently, explaining the reduced star formation rates of these galaxies \citep[see also][]{davies20,davies22}.

\subsubsection{Two modes self-regulation in galaxies with massive SMBHs}
\label{sec:newmodes}

In Section \ref{sec:secularfb} we outlined the picture described by \citet{bower17} and \citet{mcalpine18}, in which EAGLE's BHs predominantly grow in two phases separated by a transitional non-linear growth phase. \citet{mcalpine18} name the final, post-NLG phase the ``AGN feedback regulated phase'', implying that the regulation of BH and galaxy growth is dominated by AGN feedback, but our findings show that this is not necessarily the case. EAGLE galaxies can remain star-forming and regulated by a mixture of stellar and low-level AGN feedback long after the NLG phase, so long as no mergers disrupt the gas disc. We therefore propose that the post-NLG phase can be further split into two modes:
\begin{itemize}
    \item {\bf Stellar \& AGN feedback co-regulation:} The presence of a co-rotating gas disc in the galaxy largely decouples the growth of the BH from the properties of the host halo. Gas that cools from the CGM settles into the disc and the BH need only grow enough (and inject enough AGN feedback) to regulate the gas that migrates to the centre of the disc. The CGM is minimally affected by this AGN feedback and the galaxy remains star-forming. The BH may be {\it locally} self-regulating with AGN feedback, but the galaxy and its halo are regulated by a mixture of stellar and low-level AGN feedback.
    
    \item {\bf AGN feedback regulation:} When the galaxy disc is disrupted/destroyed, the gas cooling from the CGM (in addition to the cold gas already in the galaxy) can reach the BH's vicinity. The BH must therefore regulate inflow from the halo; this involves ejecting a significant fraction of the CGM in order to reduce the density, and hence the cooling rate, of the remaining circumgalactic gas. This regulation mode requires the injection of far more AGN feedback and the growth of the BH to a higher mass. The amount of energy required may depend on halo properties such as the binding energy. The prevention of cooling from the CGM reduces or quenches star formation in the galaxy, and AGN feedback becomes the dominant energy injection mechanism.
\end{itemize}

The BHs in the \secular{} and \earlysecular{} galaxies remain in the first of these modes after they begin to regulate their own growth; changes to the halo's formation time and binding energy therefore have little influence on the energy injected by AGN (and in turn the halo $f_{\rm b}$ and galaxy sSFR), as shown in Section \ref{sec:results:assembly}. In the \merger{} and \earlymerger{} systems, however, a merger occurs that induces a transition to the second of these modes, transforming the subsequent evolution of the BH, CGM and galaxy, and recoupling the properties of the central BH to the properties of the halo. The transition between these two modes is likely essential to the diversity in BH, CGM and galaxy properties at fixed halo mass in the EAGLE galaxy population.

\subsection{Re-interpreting correlations in the EAGLE population}
\label{sec:results:correlations}

Previous studies of the galaxy populations in the large-volume EAGLE (and IllustrisTNG) simulations have concluded that diversity in the properties of BHs, galaxies and their gaseous haloes at fixed halo mass stems from differences in halo binding energy, which in turn is assumed to be the result of differences in assembly/collapse time. The basis for these conclusions was the clear correlation between the inner halo binding energy, $E_{2500}^{\rm DMO}$, and $M_{\rm BH}$ at fixed $M_{200}$. The explanation for this correlation came from self-regulation arguments; tightly-bound haloes more effectively confine gas at the halo centre, requiring the injection of more AGN feedback (and hence a higher $M_{\rm BH}$) to expel gas and regulate the growth of the BH and galaxy \citep[e.g.][]{boothschaye10,boothschaye11,davies19,davies20,davies21}. However, this explanation neglects the influence of the galaxy disc and the role of merger events, which we have shown to be crucial to the evolution of the BH and its host galaxy in this study. We therefore now revisit this correlation and consider its origin in light of this new information.

\begin{figure}
\includegraphics[width=\columnwidth]{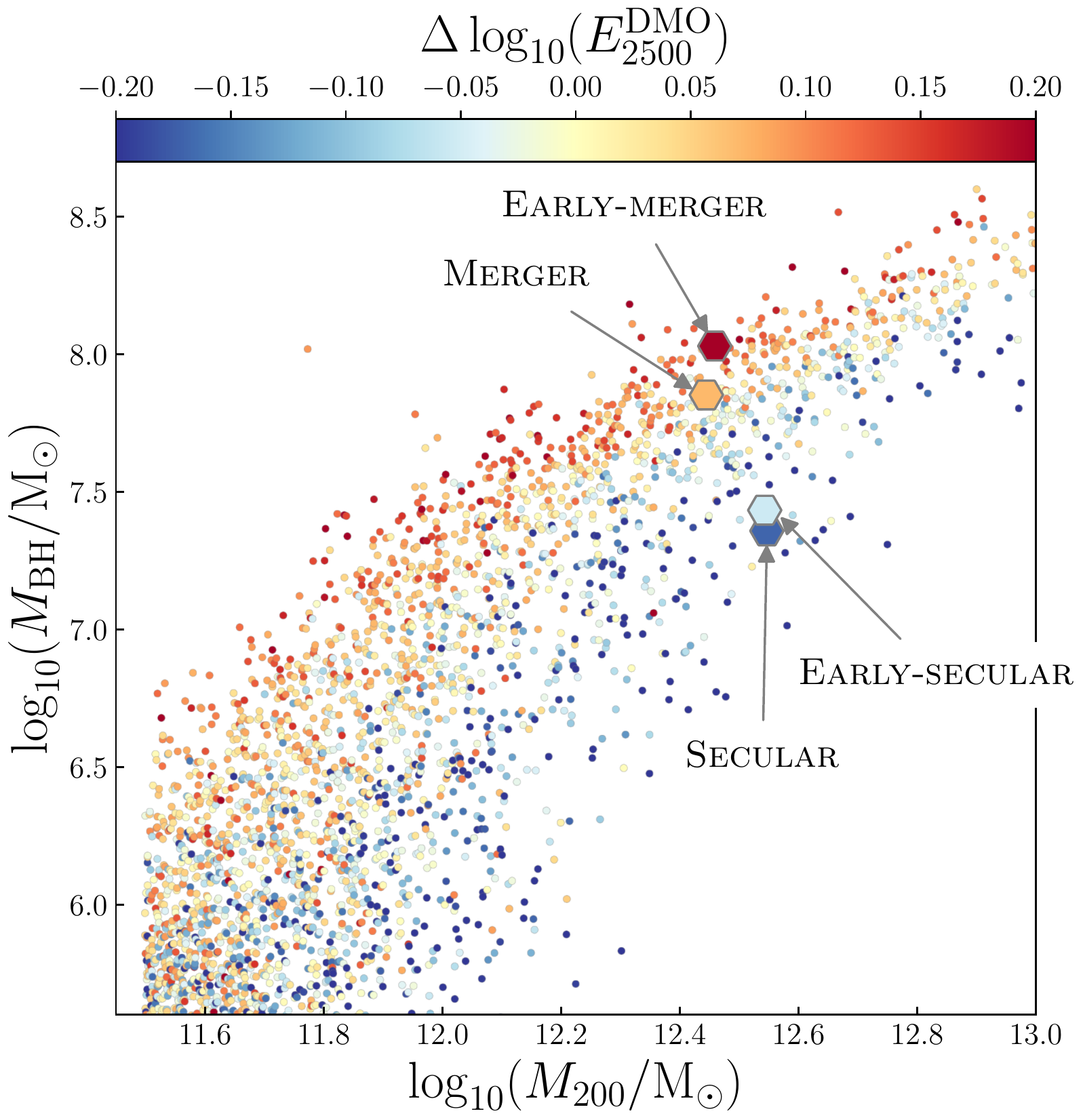}
\vspace{-4mm}
\caption{The black hole mass, $M_{\rm BH}$, as a function of $M_{200}$ for the galaxy population in the Ref-L100N1504 EAGLE simulation. Datapoints are coloured according to the residuals of the $\log_{10}(E_{2500}^{\rm DMO})-\log_{10}(M_{200}/{\rm M}_\odot)$ relation, where $E_{2500}^{\rm DMO}$ is the binding energy within a sphere enclosing 2500 times the critical density for each halo's counterpart in a collisionless dark matter simulation. At fixed $M_{200}$, haloes that are intrinsically more tightly-bound host more massive black holes. Our genetically-modified systems are overlaid; our modifications to their assembly histories cause them to span the scatter in the population. Adjusting the assembly time in the absence of mergers changes $E_{2500}^{\rm DMO}$ but not $M_{\rm BH}$, while mergers can both drive high $M_{\rm BH}$ and cause haloes to be more intrinsically tightly-bound. Merger events appear to be required for establishing both diversity in $M_{\rm BH}$ and a correlation with the halo binding energy at fixed $M_{200}$.}
\vspace{-4mm}
\label{fig:energy}
\end{figure}

We show the correlation for the galaxy population in the EAGLE Ref-L100N1504 simulation in Figure \ref{fig:energy}, plotting $M_{\rm BH}$ as a function of $M_{200}$ for all central galaxies with $M_{200}>10^{11.5}$ \msol{}, and colouring the data points by the residuals (in log space) of the $E_{2500}^{\rm DMO}-M_{200}$ relation\footnote{The residuals are taken with respect to a running median value obtained through the locally weighted scatterplot smoothing method \citep[LOWESS, e.g.][]{cleveland79}.}. The marker colours reveal a very strong positive correlation between $E_{2500}^{\rm DMO}$ and $M_{\rm BH}$ at fixed $M_{200}$. The locations of our genetically-modified systems are overlaid with larger symbols, coloured in the same fashion as the wider population; for each system, we calculate $E_{2500}^{\rm DMO}$ and find the residual from the population median at the system's halo mass. 

The modified systems span the majority of the scatter in $M_{\rm BH}$ at $M_{200}\approx 10^{12.5}$ \msol{}, and their $E_{2500}^{\rm DMO}$ values agree well with the underlying correlation. The systems that experienced merger events not only host overmassive BHs but also have high inner halo binding energies, and vice-versa. The \merger{} and \secular{} systems have very similar assembly times and differ only by the presence, or lack of, a major merger, yet the \merger{} halo is significantly more tightly bound for its mass. The inner halo binding energy is therefore not only set by the assembly and collapse time, but can be strongly increased by individual mergers. Comparing the \secular{} and \earlysecular{} systems shows that earlier collapse does yield a higher $E_{2500}^{\rm DMO}$, but the increase is far smaller than that caused by a major merger, and it does not cause enhanced BH growth.

This result aligns well with the early predictions of \citet{neto07}, who demonstrated that the connection between halo concentration and formation time is clearer when the definition of formation time includes all of a halo's major progenitors and not just one, indicating that mergers play a key role in setting the concentration (and hence the binding energy). More recently, \citet{rey19} used genetic modification to increase the variance of the overdensity field in a halo's initial conditions, increasing the number of significant mergers in the merger tree; this change also caused the halo concentration to increase, providing further evidence for this connection.

Our findings suggest that the correlation between the halo binding energy and the BH mass emerges because mergers help to grow BHs to high masses {\it and} can significantly increase the binding energy of the host dark matter halo. This picture also provides an explanation for why earlier-assembling haloes in cosmological simulations tend to have higher binding energies and host quenched central galaxies with overmassive central SMBHs \citep[e.g.][]{matthee19,monterodorta20}; these systems reside in denser, more clustered environments \citep{sheth04,gao05} and are therefore likely to experience more mergers throughout their evolution.

\section{Summary and Discussion}
\label{sec:summary}

In this study, we have used the genetic modification technique \citep{roth16} to assess how supermassive black hole (BH) growth and the impact of AGN feedback are influenced by individual galaxy-galaxy merger events and the overall assembly history of the host halo. This experiment was motivated by predictions from cosmological simulations (see references in Section \ref{sec:intro}) that mergers can induce BH growth and AGN feedback, and that the BH mass correlates strongly and positively with the binding energy of the dark matter halo, which in turn is correlated with the overall assembly time.

We performed zoom simulations of a star-forming disc galaxy of stellar mass $M_\star=4.3\times 10^{10}$ \msol{} and host halo mass $M_{200}=3.4\times 10^{12}$ \msol{} with the recalibrated high-resolution version of the EAGLE galaxy formation model \citep{schaye15,crain15}. Using the initial conditions generator \textsc{genetIC} \citep{stopyra20}, we modified the initial conditions of our fiducial galaxy and its host halo to generate four new variants of it with systematically adjusted assembly histories, designed such that we could independently assess the roles of mergers and the overall assembly time in a controlled galaxy formation experiment.

First, we produced two modified variants of our galaxy for which no significant mergers (i.e. where the stellar mass ratio $\mu>0.1$) occur after the host halo reaches the critical mass, \Mcrit{}, above which BHs are able to grow via accretion in the EAGLE model \citep[see][]{bower17,mcalpine18}. One of these systems (\secular{}) has a similar assembly time to the unmodified case, while the other (\earlysecular{}) assembles earlier. This earlier assembly time causes the inner dark matter halo of the \earlysecular{} case to be intrinsically more tightly-bound; when measured in an equivalent dark matter-only simulation, its present-day binding energy, $E_{2500}^{\rm DMO}$, is 30\% higher than the \secular{} case.

Comparing these systems allowed us to isolate how differences in the halo assembly time influence the growth of the central BH, free of the influence of significant mergers. We find that these differences do not drive significant changes to the present-day properties of the galaxy and its host halo. The \secular{} and \earlysecular{} galaxies both reach the same present-day stellar mass and remain on the star-forming main sequence (Figure \ref{fig:secularmass}), and their central BHs grow to approximately the same mass and inject the same amount of AGN feedback energy, both in absolute terms and as a fraction of the total feedback energy, and their gaseous haloes remain baryon-rich (Figure \ref{fig:secularenergy}).

The BHs grow according to a three-phase process, as is expected in the EAGLE model \citep{mcalpine18}; they initially remain close to the seed mass, undergo non-linear growth (NLG) when the halo mass exceeds \Mcrit{}, and then grow more steadily once AGN feedback is able to expel gas from the BH vicinity. We find that differences in assembly time influence when the NLG phase occurs and the typical accretion rate in this phase, but do not change the total energy injected or the impact of the feedback on the galaxy-CGM ecosystem. Once the NLG phase ends, the growth of the BH and galaxy are regulated by a combination of stellar and low-level AGN feedback, which contribute approximately 70\% and 30\% of the feedback energy injected by $z=0$ respectively.

To test the influence of mergers, we produced two more sets of initial conditions, \merger{} and \earlymerger{}, designed to yield haloes with similar assembly times to the \secular{} and \earlysecular{} galaxies respectively, but also experience a major merger ($\mu=0.46$) after crossing \Mcrit{}. At $z=0$, we find that these systems have approximately the same halo and stellar masses as their secularly-evolving counterparts, but are either quenched or reside in the green valley (Figure \ref{fig:mergermass}). The mergers cause the BHs in these systems to grow significantly more massive than the BHs in their secularly-evolving counterparts, and inject significantly more AGN feedback energy (by factors of $\approx 3$ and $\approx 4$ respectively). AGN dominate the feedback energy injected into these systems, contributing $\approx 55\%$ and $\approx 59\%$ of the total integrated energy respectively. The expulsive nature of this feedback depletes the haloes of their baryons, and they retain only $\approx 30\%$ of the cosmic baryon fraction by the present day (Figure \ref{fig:mergerenergy}).

We attribute these results to the vital importance of the galaxy disc in determining the conditions in the vicinity of the central BH. The \secular{} and \earlysecular{} galaxies retain strong co-rotating gas discs throughout their evolution, with $80-90\%$ of the kinetic energy of the gas invested in co-rotational motion. This rotational support reduces the inflow rate towards the BH, suppressing its growth and reducing the feedback energy required to maintain self-regulation. The BHs in these systems appear to undergo non-linear growth to a minimum mass at which AGN feedback can expel gas from their vicinity, and then grow very little thereafter. The majority of the gas cooling from the halo settles onto the disc and fuels continued star formation, with a small minority fuelling slow growth of the BH. The disc therefore decouples the growth of the BH from the properties of the host halo, and hence changes in assembly time and binding energy have little influence on our secularly-evolving galaxies.

In our \merger{} and \earlymerger{} systems, the disc is disrupted, allowing gas in the galaxy and halo to fuel BH growth. This increases the AGN feedback energy required to maintain self-regulation, and causes a transformation of the CGM and the quenching of star formation. The disruption of the disc (through mergers or otherwise) therefore appears to be key to the establishment of diversity in the masses of BHs, and to coupling the properties of BHs with those of their host haloes. 

To summarise, we have extended the commonly-adopted picture that the baryon cycle and (internal) quenching is almost entirely dominated by stellar/SN feedback below a threshold halo mass $M_{200}\simeq10^{12}$ M$_{\odot}$, and by AGN feedback above it, by demonstrating unambiguously that the dominant feedback mechanism is set by both the halo properties and the merger history. Our experiment suggests that the halo must exceed this mass threshold {\it and} a disruptive event must occur to yield an overmassive BH and a quenched galaxy. We propose that once galaxies host massive central BHs, their growth can be regulated in one of two modes: (i) co-regulation by stellar feedback and low-level AGN feedback in secularly-evolving, star-forming disc galaxies, and (ii) AGN-dominated regulation in systems without discs, which are likely quenched due to the influence of integrated AGN feedback on the CGM. At high $M_{200}$, AGN feedback is not guaranteed to dominate and quenching need not occur, explaining why galaxies continue to form stars in haloes far exceeding $M_{200}\simeq10^{12}$ M$_{\odot}$ in EAGLE. The transition of galaxies between these two modes may be essential to the diversity of galaxy properties seen at this mass scale.

We concluded by reconsidering the origin of strong positive correlations between the BH mass and the host halo binding energy seen at fixed halo mass in the EAGLE population, in light of our results. We placed the properties of our modified systems in the context of this correlation (Figure \ref{fig:energy}), and deduced that the correlation likely emerges because mergers are key to growing BHs to high masses {\it and} they can significantly increase the binding energy of the underlying dark matter halo. The controlled nature of our genetic modification experiment was essential for developing this understanding of how the properties, assembly history and merger history of dark matter haloes collectively influence the evolution of their central galaxies and BHs.

While the behaviour of the single galaxy that we have simulated is not guaranteed to be universal in the galaxy populations of the EAGLE cosmological simulation volumes, circumstantial evidence for the essential role of mergers does exist in the EAGLE Ref-L100N1504 simulation. Galaxies residing in $\sim L^\star$ haloes that have far exceeded \Mcrit{} but host undermassive BHs and remain CGM-rich tend to have rotation-dominated stellar kinematics, while those with overmassive BHs and gas-poor haloes tend to have dispersion-dominated kinematics indicative of disruptive mergers in their past \citep{davies20}.

To more firmly establish a connection between mergers and the impact of AGN within the population, one could compare the properties of two samples of galaxies with similar present-day halo masses, but where galaxies in one sample have experienced a disruptive merger when $M_{200}>$\Mcrit{} and those in the other have not. Assembling such samples presents a challenge, however, primarily due to difficulties in defining what makes a merger `disruptive'. Whilst a high stellar mass ratio is likely a good indicator, other factors such as the morphologies of the merging galaxies, gas-richness, orbital configuration (i.e. prograde vs. retrograde) or orbit type (i.e. ``spiral-in'' or ``head-on'') may be equally important, with even 1:1 mergers often causing little disruption if they are gas-rich/prograde/spiral-in \citep[e.g.][]{hopkins09,font17,garrisonkimmel18,martin18,peschken20,zeng21}.

In our experiment we were able to systematically adjust the mass ratio of mergers to make them more or less disruptive, but we did not control for any of the above extra characteristics of mergers. Some characteristics are influenced by each other; for example, when we increase the mass ratio, we find that mergers become more ``head-on'' with smaller impact parameters. We expect that in future work, we will be able to identify the most important characteristics of mergers by genetically-modifying the angular momentum of a galaxy and its progenitors, allowing for controlled adjustment of a merger's trajectory \citep{cadiou21,cadiou22}.

We have shown in this work that galaxies which retain strong co-rotating gas discs can remain star-forming and minimally influenced by AGN feedback at halo masses above the threshold at which AGN are expected to dominate and transform the system. We have focused here on the $\sim L^\star$ mass scale in order to explain diversity in the properties of Milky Way-like galaxies, however this picture may also apply at much higher masses and in more extreme systems. While the likelihood of a galaxy experiencing a disruptive merger increases with stellar mass \citep{martin18}, a rare population of very massive ($M_\star>10^{11.3}$ \msol{}) blue ``super-spiral'' galaxies has been observed at low redshift \citep{ogle19}. These galaxies exhibit a mixture of old and young stellar populations characteristic of early assembly and a consistently high star formation rate; they may therefore represent an extreme, high-mass case of the behaviour seen in our \earlysecular{} system. It should be possible to test this hypothesis using a genetic modification experiment, in which the assembly of a massive halo ($M_{200}\approx 10^{13}$ \msol{}) hosting a quenched spheroidal galaxy is accelerated and the merger history is made as secular as possible. If these modifications yield similar changes to those in our \earlysecular{} system, the central galaxy may become a super-spiral that remains star-forming until $z=0$, long after one might expect it to have been quenched by AGN feedback. 

\section*{Acknowledgements}
We thank the reviewer for their comments and suggestions, which improved this manuscript. JJD would like to thank Corentin Cadiou and the GMGalaxies team at UCL for helpful discussions and support. This study was supported by the European Union's Horizon 2020 research and innovation programme under grant agreement No. 818085 GMGalaxies. AP and RAC are supported by the Royal Society. This study used computing equipment funded by the Research Capital Investment Fund (RCIF) provided by UKRI, and partially funded by the UCL Cosmoparticle Initiative. It also made use of high performance computing facilities at Liverpool John Moores University, funded by the Royal Society and LJMU’s Faculty of Engineering and Technology. We thank the EAGLE team for making the particle data \citep{eagleparticledata} and galaxy catalogues \citep{mcalpine16} for the Ref-L100N1504 simulation publicly available. Analysis was performed in Python using \textsc{pynbody} \citep{pontzen13} and \textsc{tangos} \citep{pontzentremmel18}.

\vspace{-4mm}

\section*{Data Availability}

The data underlying this article will be shared on reasonable request to the corresponding author.

\vspace{-4mm}



\bibliographystyle{mnras}
\bibliography{bibliography}



\bsp	
\label{lastpage}
\end{document}